\documentclass{ws-mpla_arXiv12057072} 
\usepackage{graphicx}                 

\begin{document}

\title{\vspace*{-1.25cm}ON VACUUM-ENERGY DECAY FROM PARTICLE PRODUCTION}

\author{F.R. Klinkhamer}
\address{Institute for Theoretical Physics,
Karlsruhe Institute of Technology (KIT),\\
76128 Karlsruhe, Germany\\
frans.klinkhamer@kit.edu}

\maketitle
\begin{abstract}
A simplified (but consistent) description of particle-production
backreaction effects in de Sitter spacetime is given.%
\vspace*{1.25\baselineskip}\newline
Journal: \emph{Mod. Phys. Lett. A} \textbf{27}, 1250150 (2012)
\vspace*{.25\baselineskip}\newline
Preprint:  arXiv:1205.7072
\vspace*{.25\baselineskip}\newline Keywords: Quantum field theory in curved
spacetime, early universe, cosmological constant, dark energy.
\vspace*{.25\baselineskip}\newline PACS: 03.70.+k, 98.80.Cq, 98.80.Es,
95.36.+x
\end{abstract}

\maketitle

\section{Introduction}
\label{sec:Introduction}

Polyakov has argued in a recent series of
papers\cite{Polyakov2007,Polyakov2009,KrotovPolyakov2010}
that de Sitter spacetime suffers from an explosive
production of particles, if these particles
have nonvanishing self-interactions.
He did not address quantitatively the issue
of backreaction, namely, how the original de Sitter spacetime is
changed by the produced particles.
Here, we offer a few heuristic remarks on this issue, awaiting
the definitive calculation of the relevant effective parameters.
Incidentally, an extensive list of references on the
problem of vacuum-energy decay can be found in the original
papers\cite{Polyakov2007,Polyakov2009,KrotovPolyakov2010}
and a useful follow-up paper\cite{Akhmedov2011} (see also
Refs.~\refcite{ZeldovichStarobinsky1977} and \refcite{Starobinsky1979}
for two early papers).

In four-dimensional de Sitter (dS)
spacetime\cite{HawkingEllis1973,GibbonsHawking1977}
with a positive cosmological constant $\Lambda$,
infrared higher-loop quantum effects
are also expected to lead to instability
of the pure (matter-less) vacuum state.\cite{Polyakov2009}
With the macroscopic timescale being set by the inverse of the
Hubble constant $H_\text{dS}$ and $\Lambda$ being
the only macroscopic energy-density scale available,
the produced energy density of particles is given by the
following expression ($c=\hbar=1$):
\begin{subequations}\label{eq:rM-production-HdS-EPlanck}
\begin{eqnarray}\label{eq:rM-production}
\dot{\rho}_{M}\,\Big|_{\text{dS},\;\rho_{M}=0} &=&
\sqrt{3}\;\gamma\;H_\text{dS}\;\Lambda\,,\\[2mm]
\label{eq:HdS}
H_\text{dS} &\equiv&
\sqrt{\Lambda\big/\big[3\,(E_{P})^{2}\big]}\,,
\quad \Lambda>0\,,
\\[2mm]\label{eq:EPlanck}
\hspace*{-8mm}
E_{P}
&\equiv&
1/\sqrt{8\pi G_{N}}\,,\quad G_{N} >0\,,
\end{eqnarray}
\end{subequations}
where
the square-root factor on the right-hand side of
\eqref{eq:rM-production} has been added for later convenience and
the dimensionless parameter $\gamma$ is a short-hand notation
for $\gamma_{V}$
(another parameter $\gamma_{M}$ will be introduced subsequently).
The parameter $\gamma >0$ involves the scalar self-coupling constants
(see further comments below) and the overdot on
the left-hand side of \eqref{eq:rM-production} stands for differentiation
with respect to a cosmic time coordinate $t$ to be defined later (the
resulting matter perturbation breaks the original de Sitter symmetry). It
needs to be emphasized, right from the start, that Polyakov's tentative
result \eqref{eq:rM-production} solely relies on the careful study of quantum
field theory in curved spacetime and does not require the introduction of new
theories.

The constant growth rate from \eqref{eq:rM-production}
only holds as long as
$\rho_{M} \ll \Lambda$, so that backreaction effects can be neglected.
In order to address the further evolution ($\rho_{M} \sim \Lambda$),
we replace the cosmological constant $\Lambda$ by
a dynamic vacuum energy density $\rho_{V}$.
That is, we split the backreaction effects from the produced
particles into a standard-matter-type component
and a vacuum-type component.

\section{Dynamic Vacuum Energy}
\label{sec:Dynamic-vacuum-energy}

For our purpose, the following classical
action\cite{KlinkhamerVolovik2008a,KlinkhamerVolovik2008b}
can be used:%
\begin{subequations}\label{eq:action-LM}
\begin{eqnarray}\label{eq:action}
\hspace*{-0.0mm}
S_\text{class}
&=&
-\int\,d^4x\,\sqrt{-\text{det}(g)}\;
\left[ \frac{1}{2}\,(E_{P})^{2}\,R + \epsilon_{V}(q) +
\mathcal{L}_{M}\right],
\\[2mm]\label{eq:LM}
\hspace*{-0.0mm}
\mathcal{L}_{M} &=&
-\frac{1}{2}\,(\partial\phi)^2+\frac{1}{2}\,m^2\,\phi^2
+\frac{1}{4}\,\kappa\,\phi^4\,,
\end{eqnarray}
\end{subequations}
where the matter part \eqref{eq:LM} is entirely standard,
consisting of a massive real scalar field $\phi(x)$
with quartic coupling constant $\kappa> 0$.
In \eqref{eq:action}, $\epsilon_{V}(q)$ is a generic function of
a $q$--type field, which is essentially a nonfundamental scalar
field $q(x)$ with a constant nonzero expectation value $q_{0}$
in the equilibrium state.
One particular realization of $q(x)$ is
by use of a three-form gauge field $A(x)$;
see Refs.~\refcite{KlinkhamerVolovik2008a} and \refcite{KlinkhamerVolovik2008b}
for details and original references.

The vacuum energy density in the corresponding gravitational field
equation is denoted by $\rho_{V}$ and differs, in general,
from the energy density $\epsilon_{V}$ entering the action.
In terms of a single $q$--type field,
the gravitating vacuum energy density takes the following
form\cite{KlinkhamerVolovik2008a,KlinkhamerVolovik2008b}:%
\begin{equation}
\rho_{V}(q)=-P_{V}(q)=\epsilon_{V}(q)-q\; \frac{d \epsilon_{V}(q)}{d q}\,.
\end{equation}
Henceforth, we assume that $\rho_{V}$ takes sub-Planckian values
and that the scalar particles are sufficiently heavy,\cite{Polyakov2009}%
\begin{subequations}\label{eq:rhoVbounds-Mbound}
\begin{eqnarray}
\label{eq:rhoVbounds}
0 &<& \rho_{V} \ll (E_{P})^{4}
\approx (2.44\times 10^{18}\:\text{GeV})^{4}\,,
\\[2mm]
\label{eq:Mbound}
m &\gg& \sqrt{\rho_{V}/(E_{P})^{2}}\,.
\end{eqnarray}
\end{subequations}
Condition \eqref{eq:rhoVbounds} allows us to rely on classical gravity.
In the cosmological context to be discussed shortly,
condition \eqref{eq:Mbound} implies that the ``response-time''
of the particle wave function
is very much less than the relevant cosmological timescale.
Condition \eqref{eq:Mbound} in the cosmological context
also suggests that the produced
scalar particles are nonrelativistic, their rest mass  being
very much larger than the Gibbons--Hawking
temperature\cite{GibbonsHawking1977}
of the corresponding de Sitter spacetime,
$m\gg T_\text{GH}=H_\text{dS}/2\pi$.

The action \eqref{eq:action-LM} is
invariant under general coordinate transformations
or, more precisely, diffeomorphisms.
Diffeomorphism invariance implies energy-momentum
conservation of the matter component
(cf. Appendix~E.1 of Ref.~\refcite{Wald1984}). Here, the matter component
is given by the dynamic vacuum energy density $\rho_{V}=-P_{V}$
and the scalar field $\phi$.

\section{FRW Equations}
\label{sec:FRW-equations}

Now let us restrict ourselves to a spatially flat Robertson--Walker metric
(in standard comoving coordinates\cite{HawkingEllis1973,Wald1984} with
cosmic time $t$), a homogeneous
pressureless-perfect-fluid component
($w_{M}\equiv P_{M}/\rho_{M}=0$), and a homogeneous vacuum-energy component
($w_{V}\equiv P_{V}/\rho_{V}=-1$). Then, the Friedmann--Robertson--Walker
(FRW) equations [in dimensionless form given by \eqref{eq:ODEs-hdot} and
\eqref{eq:ODEs-friedmann} below] are consistent only when using the following
generalization of \eqref{eq:rM-production}:
\begin{subequations}\label{eq:FRW-rMdot-rVdot}
\begin{equation}\label{eq:FRW-rMdot} \dot{\rho}_{M} +
3\,H\,\rho_{M} =
\gamma\;\sqrt{|\rho_{V}|/(E_{P})^{2}}\;\rho_{V}\,,
\end{equation}
and the
following vacuum-energy-density equation:
\begin{equation}\label{eq:FRW-rVdot}
\dot{\rho}_{V} =
-\gamma\;\sqrt{|\rho_{V}|/(E_{P})^{2}}\;\rho_{V}\,,
\end{equation}
\end{subequations}
where the equation of state $P_{V}=-\rho_{V}$ has been used on the left-hand
side. The absolute value $|\rho_{V}|$ has been employed in the square-roots
on the right-hand sides of \eqref{eq:FRW-rMdot-rVdot}, but this is, strictly
speaking, not necessary if \eqref{eq:rhoVbounds} holds.

Before presenting the complete set of differential equations,
let us comment on the physical interpretation
of \eqref{eq:FRW-rMdot} and \eqref{eq:FRW-rVdot}. The
first set of comments concerns the starting point, the de Sitter-spacetime
calculation giving the right-hand sides of \eqref{eq:rM-production} and
\eqref{eq:FRW-rMdot}. Given the theory \eqref{eq:action-LM}, the
dimensionless parameter $\gamma>0$ involves expressions containing positive
powers of the quartic coupling constant $\kappa$. The main results of
Polyakov's investigations are, first, that higher-loop quantum corrections
may contribute the seed for $\gamma$ (for example, from a 1-loop Feynman
diagram implicit in the second paragraph of Sec.~5 in
Ref.~\refcite{Polyakov2009} but not evaluated explicitly) and, second, that a
chain-reaction-type evolution can amplify microscopic effects to macroscopic
ones [see Eq.~(27) in Ref.~\refcite{Polyakov2009} for an example of this
explosive behavior and Sec.~\ref{sec:Discussion} for further discussion].
Both results are infrared effects, with the whole of de Sitter spacetime
contributing.

The second set of comments concerns the backreaction. Purely from the
classical theory \eqref{eq:action-LM}, the right-hand sides of
\eqref{eq:FRW-rMdot} and \eqref{eq:FRW-rVdot} should vanish: see,
respectively, Eqs.~(4.6) and (4.5) in Ref.~\refcite{KlinkhamerVolovik2008b} for
the case of a constant gravitational coupling parameter $G=G_{N}$. In the
stationary background of the classical de Sitter spacetime and with classical
$q$--type fields entering the vacuum energy density $\rho_{V}$, the scalar
quantum field theory corresponding to \eqref{eq:LM} would give rise to
particle production from infrared quantum
effects\cite{Polyakov2009,KrotovPolyakov2010} as given by
\eqref{eq:rM-production}. In the dynamic context [time-dependent vacuum
energy density $\rho_{V}(t)$], this results in the expression shown on the
right-hand side of \eqref{eq:FRW-rMdot}, with the opposite expression on the
right-hand side of \eqref{eq:FRW-rVdot} from energy-momentum conservation
(tracing back to the diffeomorphism invariance of the theory, including
quantized-matter effects). There are further terms on the right-hand-sides of
\eqref{eq:FRW-rMdot} and \eqref{eq:FRW-rVdot}, which are of the form
$\pm\,\gamma_{M}\;\sqrt{|\rho_{V}|/(E_{P})^{2}}\;\rho_{M}$ for
$\gamma_{M}> 0$ and which correspond to a type of stimulated emission of matter
particles (cf. Sec.~5 in Ref.~\refcite{Polyakov2009}). These terms are the
outcome of the chain-reaction-type evolution mentioned in the previous
paragraph but will, for the moment, not be considered explicitly
($\gamma_{M}=0$).

As promised, we give the complete set of
ordinary differential equations (ODEs), obtained by adding the two FRW
equations which follow from the action \eqref{eq:action}
for the spatially flat ($k=0$) Robertson--Walker metric.
Introducing dimensionless variables by use of appropriate powers of
$E_{P}$, these ODEs are
\begin{subequations}\label{eq:ODEs}
\begin{eqnarray} \label{eq:ODEs-rVdot}
\hspace*{-0.0mm}
\dot{r}_{V} &=& -\gamma\;|r_{V}|^{1/2}\,r_{V}\,,
\\[2mm]
\label{eq:ODEs-rMdot} \hspace*{-0.0mm} \dot{r}_{M} + 3\,h\,r_{M}
&=&+\gamma\;|r_{V}|^{1/2}\,r_{V}\,,
\\[2mm]
\label{eq:ODEs-hdot}
\hspace*{-0.0mm}
2\,\dot{h} &=& -r_{M}\,,
\\[2mm]
\label{eq:ODEs-friedmann}
\hspace*{-0.0mm}
3\,h^{2} &=& r_{V} + r_{M} \,,
\end{eqnarray}
\end{subequations}
with the dimensionless Hubble parameter $h(\tau)$ and the
dimensionless cosmic time $\tau$ (the overdot now stands
for differentiation with respect to $\tau$).
The dimensionless variable $r_{M}(\tau)\geq 0$ corresponds
to the standard-matter energy density $\rho_{M}(t)\geq 0$ with
constant equation-of-state parameter $w_{M}= 0$.
Similarly, $r_{V}(\tau)=-p_{V}(\tau)$
corresponds to $\rho_{V}(t)=-P_{V}(t)$.
Elaborating on the previous discussion of energy-momentum conservation,
the differential system \eqref{eq:ODEs} is seen to be consistent:
\eqref{eq:ODEs-hdot} follows from
taking the time derivative of \eqref{eq:ODEs-friedmann}
and using \eqref{eq:ODEs-rVdot} and \eqref{eq:ODEs-rMdot}.

In order to match the setup of \eqref{eq:rM-production},
the boundary conditions for the ODEs
\eqref{eq:ODEs-rVdot}, \eqref{eq:ODEs-rMdot}, and \eqref{eq:ODEs-hdot}
are taken as follows:
\begin{subequations}\label{eq:BCS}
\begin{eqnarray}
\label{eq:BCS-rV}
\hspace*{-0.0mm}
r_{V}(1) &=& \lambda >0\,,
\\[2mm]
\label{eq:BCS-rM}
\hspace*{-0.0mm}
r_{M}(1)  &=& 0\,,
\\[2mm]
\label{eq:BCS-h}
\hspace*{-0.0mm}
h(1) &=& \sqrt{\lambda/3}\,,
\end{eqnarray}
\end{subequations}
with the Friedmann equation \eqref{eq:ODEs-friedmann}
acting as a constraint. The boundary conditions are, therefore,
characterized by the single
number $\lambda\equiv\Lambda/(E_{P})^4$ and the ODEs
\eqref{eq:ODEs} by the single model-parameter $\gamma$.

\section{Exact $\boldsymbol{k=0}$ FRW Solution}
\label{sec:Exact-FRW-solution}

\begin{figure*}[t]
\vspace*{-0mm}
\begin{center}
\hspace*{-5mm}                           
\includegraphics[width=1.06\textwidth]{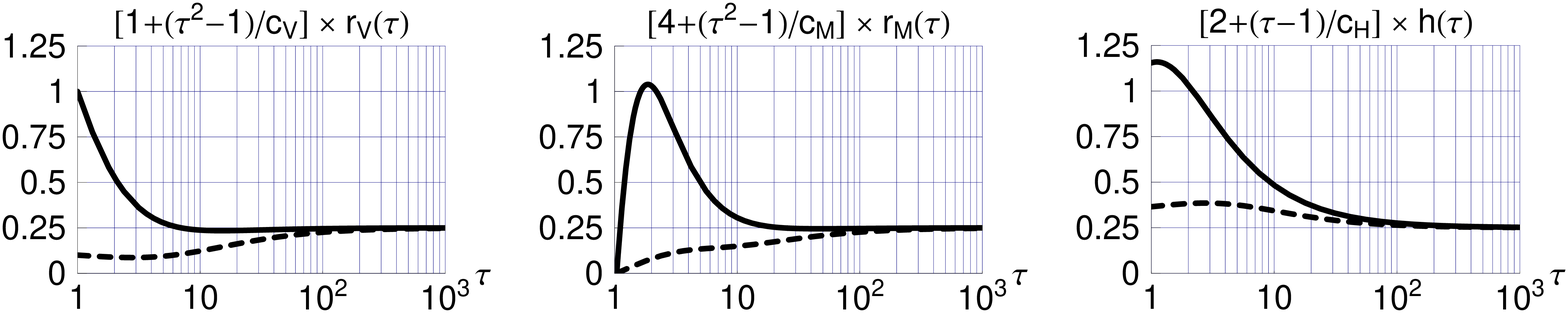}
\end{center}
\vspace*{-4mm} \caption{Numerical solution of the
ODEs \eqref{eq:ODEs} for vacuum-energy decay parameter $\gamma=1$
with different boundary conditions \eqref{eq:BCS}, the upper (full) curves
corresponding to an initial dimensionless cosmological
constant $\lambda\equiv\Lambda/(E_{P})^4=1$
and the lower (dashed) curves to $\lambda=1/10$. The curves are
rescaled by constant numerical factors at $\tau=1$ and by appropriate $\tau$
monomials for $\tau\gg 1$. The coefficients $\{c_{V},\,  c_{M},\, c_{H}\}$
used in this rescaling are defined in \eqref{eq:asymp} and
take the numerical values $\{16,\,  12.28,\,6.14\}$.
The model universe
for $\tau\sim 1$ resembles a segment of de Sitter spacetime
\eqref{eq:dS-solution} with approximately constant Hubble parameter,
$h(\tau)\sim h(1)=\sqrt{\lambda/3}$.
As $\tau$ increases, the vacuum energy density
$r_{V}(\tau)$ drops from the value $\lambda$ to zero
and the model universe approaches Minkowski
spacetime with $h=0$. \vspace*{0cm}} \label{fig:1}
\end{figure*}

For $\gamma=0$, the solution of the ODEs \eqref{eq:ODEs} with boundary
conditions \eqref{eq:BCS} corresponds to the standard (eternal) de Sitter
spacetime,\cite{HawkingEllis1973}
having the following constant functions:%
\begin{eqnarray}\label{eq:dS-solution}
\Big[\big\{r_{V}(\tau), \,r_{M}(\tau),
\,h(\tau)\big\}\Big]^{(\gamma=0)} &=& \big\{\lambda,\,  0,\,
\sqrt{\lambda/3}\,\big\}\,.
\end{eqnarray}
For $\gamma\to\infty$, the solution
rapidly approaches the standard matter-dominated FRW solution (see the
asymptotic result below). For $\gamma=1$, the numerical
solution is shown in Fig.~\ref{fig:1}.

Given the relative simplicity of the differential system
\eqref{eq:ODEs} and boundary conditions \eqref{eq:BCS}, it is also
possible to obtain the exact solution for $\gamma>0$,
\begin{subequations}\label{eq:exact-solution}
\begin{eqnarray}
\label{eq:exact-rV}
\Big[r_{V}(\tau)\Big]^{(\gamma>0)} &=&
\left( \frac{1}{1/\sqrt{\lambda}+\gamma\,(\tau-1)/2}\right)^2\,,
\\[2mm]
\label{eq:exact-rM}
\Big[r_{M}(\tau)\Big]^{(\gamma>0)} &=&
3\;\left(\Big[h(\tau)\Big]^{(\gamma>0)}\right)^2
-  \Big[r_{V}(\tau)\Big]^{(\gamma>0)}\,,
\\[2mm]
\label{eq:exact-h}
\Big[h(\tau)\Big]^{(\gamma>0)} &=&
\frac{1}{\kappa +\theta}
\nonumber\\
&&\times
\frac{
(2/\gamma)\,
\big[
\left(\kappa+\theta\right)^\eta - \kappa^\eta
\big] +
\sqrt{1/3}\,
\big[
\left( \eta + 1 \right) \, \left( \kappa + \theta \right)^\eta
+
\left(\eta -1\right)\,\kappa^\eta
\big]}
{\;\;\,\sqrt{3}\;\,
\big[
\left( \kappa + \theta \right)^\eta -\kappa^\eta
\big]
+\;(\gamma/2)\,
\big[
\left(\eta -1\right) \,\left( \kappa + \theta \right)^\eta
+
\left( \eta + 1 \right)\,\kappa^\eta
\big]}\,,
\nonumber\\
&&
\end{eqnarray}
with definitions
\begin{equation}
\theta \equiv \tau-1\,,
\quad
\eta \equiv       \sqrt{1+12/\gamma^2}\,,
\quad
\kappa \equiv 2\big/\big(\gamma\,\sqrt{\lambda}\big)\,.
\end{equation}
\end{subequations}
Incidentally, the $h$ solution \eqref{eq:exact-h} results from a
generalized Riccati equation, namely, $2\,\dot{h} +3\,h^{2} = r_{V}$
with $r_{V}$ from \eqref{eq:exact-rV}.
[For the case of relativistic matter ($w_{M}=1/3$)
and the same $r_{V}$ from \eqref{eq:exact-rV}, the relevant
generalized  Riccati equation is $\dot{h} +2\,h^{2} =(2/3)\, r_{V}$,
which can also be solved explicitly.]
Remark that, for fixed $\lambda>0$ and $\tau\geq 1$, solution
\eqref{eq:exact-solution} reduces to \eqref{eq:dS-solution} as
$\gamma$ approaches $0$ from above.

The vacuum-energy density \eqref{eq:exact-rV} is seen to
drop to zero monotonically as $\tau$ runs from $1$ to infinity
(the left panel of Fig.~\ref{fig:1} shows a rescaled $r_{V}$).
The asymptotic (attractor) solution
for $\gamma>0$ has, moreover,
an equal time-dependence for the vacuum and matter components:%
\begin{subequations}\label{eq:asymp}
\begin{eqnarray}
\label{eq:asymp-rV} \hspace*{-0.0mm}
\Big[r_{V,\,\text{asymp}}(\tau)\Big]^{(\gamma>0)} &=&
\frac{1}{4}\,c_{V}\,\tau^{-2}\equiv (4/\gamma^2)\,\tau^{-2}\,,
\\[2mm]
\label{eq:asymp-rM} \hspace*{-0.0mm}
\Big[r_{M,\,\text{asymp}}(\tau)\Big]^{(\gamma>0)}  &=&
\frac{1}{4}\,c_{M}\,\tau^{-2}\equiv
\frac{2}{3}\,\Big(1+\sqrt{1+12/\gamma^2}\Big)\,\tau^{-2}\,,
\\[2mm]\label{eq:asymp-h}
\hspace*{-0.0mm} \Big[h_\text{asymp}(\tau)\Big]^{(\gamma>0)} &=&
\frac{1}{4}\,c_{H}\,\tau^{-1}\equiv
\frac{1}{3}\,\Big(1+\sqrt{1+12/\gamma^2}\Big)\,\tau^{-1}\,.
\end{eqnarray}
\end{subequations}
The asymptotic energy-density ratio depends
only on the decay parameter $\gamma$:%
\begin{equation}\label{eq:asymp-ratio}
\lim_{\tau\to\infty}\;r_{M}/r_{V} =
\frac{1}{6}\,\Big(\gamma^2+\gamma\,\sqrt{\gamma^2+12}\Big),
\end{equation}
which goes as $(1/3)\;\gamma^2$ for $\gamma \gg 1$
and as $\sqrt{1/3}\;\gamma$ for $0\leq \gamma \ll 1$.
For $\gamma=\text{O}(1)$, the ratio
\eqref{eq:asymp-ratio} is of order $1$.

\section{Discussion}
\label{sec:Discussion}

Two generalizations can be mentioned at this point. First, similar results
for the asymptotic behavior are obtained if stimulated-emission-type terms
$\mp\,\gamma_{M}\;|r_{V}|^{1/2}\;r_{M}$ are added to the right-hand sides of the
ODEs \eqref{eq:ODEs-rVdot} and \eqref{eq:ODEs-rMdot}, but, here, we keep
$\gamma_{M}=0$ for simplicity.

Second, it is also possible to consider the
de Sitter perturbation \eqref{eq:FRW-rMdot-rVdot} to hold for the
spatially-closed ($k=1$) Robertson--Walker metric.\cite{HawkingEllis1973}
The ODEs \eqref{eq:ODEs} are changed as follows:
the left-hand side of \eqref{eq:ODEs-hdot} picks
up a term $-2\,k/a^2$ and the left-hand side of \eqref{eq:ODEs-friedmann}
a term $+3\,k/a^2$, where $a=a(\tau)$ is the cosmic scale factor with Hubble
parameter $h \equiv \dot{a}/a$.
Numerical $k=1$ solutions have been obtained for the
following boundary conditions at the ``waist'' of de Sitter spacetime
(minimum distance between geodesic normals;
cf. Fig.~16 in Ref.~\refcite{HawkingEllis1973}):
$r_{V}(0) = \lambda>0$,
$r_{M}(0)  = h(0) = 0$, and $a(0)=\sqrt{3/\lambda}$.
(See Appendix~A  for corresponding analytic results.)
The behavior found numerically
is different for $\gamma$ below or above a critical
value $\gamma_{c}$ which depends on $\lambda$
[for example, $\gamma_{c}\approx 0.9215$ for $\lambda=1$].
For $\gamma < \gamma_{c}$,
the $k=1$ numerical solution has the same asymptotics as the
$k=0$ solution given by \eqref{eq:asymp}
[having, in particular, the same value for the
energy-density ratio \eqref{eq:asymp-ratio}],
even though different regions of spacetime are covered.
But, for $\gamma \geq \gamma_{c}$, the $k=1$ perturbation
\eqref{eq:FRW-rMdot-rVdot} affects the de Sitter universe
in a fundamentally different way:
a big-crunch singularity is  produced in a finite amount of time,
just as happens for the standard matter-dominated $k=1$ FRW universe.
The underlying microscopic physics determines which of
the two types of perturbations
is more appropriate, $k=0$ or $k=1$, and with which value of the
decay constant $\gamma$.

Let us return to the original theory \eqref{eq:ODEs} with $\gamma>0$,
$\gamma_{M}=0$, and $k=0$.
Considering the vacuum energy density $r_{V}(\tau)$, it is
then possible to estimate the cross-over time between the de Sitter-type
behavior \eqref{eq:dS-solution} and the FRW-type behavior
\eqref{eq:asymp-rV}.
More specifically, the  \mbox{``half-life''} of the de Sitter
universe [defined as the time needed to reduce an initial value
$r_V=\lambda$ from \eqref{eq:BCS} by a factor $1/2$]
is found to have the following
parametric behavior from \eqref{eq:exact-rV}:
\begin{equation}\label{eq:dS-half-life} t_\text{\,dS-half-life} =
2\,\big(\sqrt{2}-1\big)\;
\frac{1}{\gamma}\;\frac{E_{P}}{\sqrt{\Lambda}}\;,
\end{equation}
where the
dimensions have been restored and the right-hand side is, as expected from
\eqref{eq:rM-production}, proportional to the inverse of the Hubble constant
\eqref{eq:HdS}. Remark that
our use of the term ``half-life'' is not intended to imply an exponential
decay, which is indeed not the case for \eqref{eq:asymp-rV}.

The half-life \eqref{eq:dS-half-life} is a direct manifestation of
backreaction effects from the original de Sitter-spacetime
particle-production \eqref{eq:rM-production}. For $\gamma=\text{O}(1)$, this
backreaction timescale would be of order $1/H_\text{dS} \sim
E_{P}/\Lambda^{1/2}$. Such a timescale would be very much less than
the timescale found
previously,\cite{ZeldovichStarobinsky1977,Starobinsky1979}
which is of the order
$\big[(E_{P})^4/\Lambda\big]^{n/2}\;1/H_\text{dS}$ for $n\sim 1-2$
[recall the assumption that $\Lambda$ (or $\rho_{V}$) takes a value far below
$(E_{P})^{4}$]. Physically, the short backreaction timescale
\eqref{eq:dS-half-life} for $\gamma=\text{O}(1)$ would be due to the
chain-reaction effect mentioned in Sec.~\ref{sec:FRW-equations}. This
timescale of order $1/H_\text{dS}$
is indeed seen for the explicit solution
of the simplified kinetic equation (27) in Ref.~\refcite{Polyakov2009},
assuming the relevant overlap integrals to be of \mbox{order $1$.}
(See also the respective Sections~5 and 6 of Ref.~\refcite{Akhmedov2011}
for other examples without and with chain-reaction effects.)

Let us close with some further speculations.
Assume that Polyakov's mechanism \eqref{eq:rM-production}
is relevant not only for the
very early universe ($T \sim E_{P}$)
 but also for the present epoch ($T \sim 3\,\text{K}$).
Then, purely phenomenologically, consider having a time-dependent coupling
$\gamma(\tau)$ in the differential system \eqref{eq:ODEs}. If $\gamma(t)$
is like a step-function which drops to zero at a relatively recent moment
($t=t_\text{freeze}$) in the history of the Universe, this would fix
$\rho_{V}(t)$ at later times to the constant value
$\rho_{V}(t_\text{freeze})$. Given that the presently observed value of
$\rho_{M}/\rho_{V}$ is approximately equal to $1/3$, this would suggest that
the pre-freeze value of $\gamma$ should have been of order $1$.

But it is also possible that the parameter $\gamma$ entering
\eqref{eq:rM-production} and \eqref{eq:FRW-rMdot-rVdot} is significantly
smaller than $1$ and that the matter of the present Universe has a different
origin (not solely the product of vacuum-energy decay).

Leaving these
speculations aside, the clear priority at this moment is the
de Sitter-spacetime calculation of the effective vacuum-energy decay
parameters \mbox{$\gamma_{V}\equiv\gamma$}, $\gamma_{M}$,
and others if present.

\begin{appendix}
\section{Series-type $\boldsymbol{k=1}$ FRW Solution}
\label{app:Series-for-k=1-FRW-solution}

The basic ODEs for perturbation \eqref{eq:FRW-rMdot-rVdot}
with $\gamma>0$ in a $k=1$ Robertson--Walker metric are%
\begin{subequations}\label{eq:ODEs-k=1}
\begin{eqnarray} \label{eq:ODEs-rVdot-k=1}
\hspace*{-0.0mm}
\dot{r}_{V} +\gamma\;|r_{V}|^{1/2}\,r_{V}&=& 0\,,
\\[2mm]
\label{eq:ODEs-friedmann-k=1}
\hspace*{-0.0mm}
3\,\left(\dot{a}/a\right)^{2}+3\,k/a^{2}&=&r_{V} + r_{M} \,,
\\[2mm]
\label{eq:ODEs-addot-k=1}
\hspace*{-0.0mm}
2\,\ddot{a}/a+\left(\dot{a}/a\right)^{2}+k/a^{2}&=&r_{V}\,.
\end{eqnarray}
\end{subequations}
As discussed in Sec.~\ref{sec:Discussion},
the boundary conditions can be taken as follows:
\begin{eqnarray}\label{eq:BCS-k=1}
r_{V}(0) &=& \lambda >0\,,\quad
a(0) = \sqrt{3/\lambda}\,,\quad
\dot{a}(0) = r_{M}(0)=0\,.
\end{eqnarray}
With these  boundary conditions, it is possible
to study quantitatively the
backreaction effects from vacuum-energy decay,
for example, by imagining that $\gamma$ would
be turned off for $\tau<0$ (corresponding to the
standard de Sitter universe)
and turned on for $\tau\geq 0$
(giving the perturbed de Sitter universe).
Other boundary conditions give similar results.

Just as for the $k=0$ case in Sec.~\ref{sec:Exact-FRW-solution},
the solution of \eqref{eq:ODEs-rVdot-k=1}
with boundary condition from \eqref{eq:BCS-k=1}
can be obtained immediately,
\begin{eqnarray}
\label{eq:exact-rV-k=1}
r_{V}(\tau)&=&
\lambda\;\left( \frac{1}{1+(\gamma/2)\,(\sqrt{\lambda}\,\tau)}\right)^2\,.
\end{eqnarray}
With this $r_{V}(\tau)$,
it remains to solve the single second-order ODE
\eqref{eq:ODEs-addot-k=1}.

Given that our interest is primarily in vacuum-energy decay
for relatively small values of the decay constant
$\gamma$ and cosmological constant  $\lambda$,
the following \textit{Ansatz} turns out to be useful
for small enough positive values of the cosmic time $\tau$:
\begin{eqnarray}
\label{eq:exact-a-k=1}
a(\tau) &=&
\sqrt{3/\lambda}\;\cosh\big(\sqrt{\lambda/3}\;\tau\big)
\,\left[1+\gamma\,\sum_{n=2}^{\infty}\,c_{n}(\gamma)\,
\big(\sqrt{\lambda}\,\tau\big)^{n}\right]\,,
\end{eqnarray}
where coefficient $c_{n}(\gamma)$ is a polynomial in $\gamma$.
Inserting \textit{Ansatz} \eqref{eq:exact-a-k=1}
into the ODE \eqref{eq:ODEs-addot-k=1}
gives the following Taylor expansion at $\tau=0$:%
\begin{eqnarray}
\label{eq:k=1}
0 &=&
f_{0}(c_{2})+
f_{1}(c_{3},\,c_{2})\,\tau^{1}+
f_{2}(c_{4},\,c_{3},\,c_{2})\,\tau^{2}+ \cdots \,,
\end{eqnarray}
where the functions $f_{n}$ shown
are linear in the leading coefficient $c_{n+2}$.
It is, then, possible to solve sequentially for the
coefficients $c_{n}$. The first eight coefficients are
\begin{subequations}\label{eq:cn-k=1}
\begin{eqnarray}
\hspace*{-0.0mm}
c_{2} &=& 0\,,\quad c_{3} =-1/12 \,,\quad c_{4} = {\gamma}/32\,,
\\[2.5mm]
c_{5} &=& \left( 8 - 9\,{\gamma}^2 \right)/720 \,,
\quad
c_{6} = {\gamma}\,\left(-2+3\,{\gamma}^2 \right)/576\,,
\\[2.5mm]
c_{7} &=&
-\left( 608 - 369\,{\gamma}^2 + 810\,{\gamma}^4 \right)/362880 \,,
\\[2.5mm]
c_{8} &=&
\gamma\, \left(1288-489\,{\gamma}^2 + 1890\,{\gamma}^4 \right)/1935360\,,
\\[2.5mm]
c_{9} &=&
\left(9664 -11061\,{\gamma}^2 + 1296\,{\gamma}^4-17010\, {\gamma}^6\right)/39191040\,.
\end{eqnarray}
\end{subequations}

With $r_{V}(\tau)$ from \eqref{eq:exact-rV-k=1}
and  $a(\tau)$ from \eqref{eq:exact-a-k=1} and \eqref{eq:cn-k=1},
the function $r_{M}(\tau)$ follows directly from the
Friedmann Eq.~\eqref{eq:ODEs-friedmann-k=1}.
The resulting expression is also a series,
\begin{eqnarray}\label{eq:exact-rM-k=1}
r_{M}(\tau) &=&
\gamma\,\lambda\,\sum_{n=1}^{\infty}\,
d_{n}(\gamma)\,\big(\sqrt{\lambda}\,\tau\big)^{n}  \,,
\end{eqnarray}
where the coefficient
$d_{n}$ for $n\geq 2$ is given by an expression involving
the decay constant $\gamma$ and
the previously determined
coefficients $c_{n}$, $c_{n-1}$, $\ldots$, $c_{2}$.
Specifically, the first eight coefficients are
\begin{subequations}\label{eq:dn-k=1}
\begin{eqnarray}
d_{1} &=& 1\,, \quad d_{2} =-3\,\gamma/4 \,,
\\[2.5mm]
d_{3} &=& \left(-2 + 3\,{\gamma}^2 \right)/6\,,
\quad
d_{4} = \gamma\,\left( 6 - 5\,{\gamma}^2\right)/16\,,
\\[2.5mm]
d_{5} &=&
\left(64 - 207\,{\gamma}^2 + 135\,{\gamma}^4 \right)/720\,,
\\[2.5mm]
d_{6} &=&
-\gamma\,\left( 28 - 37\,{\gamma}^2 + 21\,{\gamma}^4 \right)/192\,,
\\[2.5mm]
d_{7} &=&
\left(-608 + 4317\,{\gamma}^2 - 3645\,{\gamma}^4 + 1890\,{\gamma}^6\right)/30240\,,
\\[2.5mm]
d_{8} &=&
\gamma\, \left(3076 - 7770\, {\gamma}^2 + 4995\, {\gamma}^4 -
          2430\, {\gamma}^6 \right)/69120\,.
\end{eqnarray}
\end{subequations}
The coefficients $d_{n}$ for $\gamma\ll 1$ and $\gamma\gg 1$
are seen to effectively give alternating series.
The same holds for coefficients $c_{n}$ from \eqref{eq:cn-k=1}.

This completes the construction
of a partial analytic solution of the ODEs \eqref{eq:ODEs-k=1},
leaving for the future
the rigorous determination of the convergence radius of
the series appearing in \eqref{eq:exact-a-k=1}
and \eqref{eq:exact-rM-k=1}.
Still, the qualitative behavior of the solution
can also be obtained numerically and has already been discussed
in the second paragraph of Sec.~\ref{sec:Discussion}.
Comparing with the numerical solutions $a_\text{num}(\tau)$
and $r_{M,\,\text{num}}(\tau)$
of the ODEs \eqref{eq:ODEs-k=1} for $\gamma \lesssim 1/10$,
the convergence of the series in \eqref{eq:exact-a-k=1}
and \eqref{eq:exact-rM-k=1} is found to be
rather poor, with a $\tau$ radius of convergence of order
$1/\sqrt{\lambda}$.

\end{appendix}

\newpage
\section*{Note Added in Proof}

In order to clarify the last sentence of Sec.~\ref{sec:Introduction},
it may be helpful to replace
the energy density $\rho_{V}$ in the rest of the article
by $\Lambda+\widetilde{\rho}_{V}$
and the dimensionless variable $r_{V}$ by $\lambda+\widetilde{r}_{V}$.
Then, $\widetilde{\rho}_{V}$ and $\rho_{M}$ correspond, respectively,
to the vacuum-type and standard-matter-type
energy density of the produced particles.
With $\gamma>0$, $k=0$, and $\widetilde{\rho}_{V}=\rho_{M}=0$ initially,
the vacuum-type component $\widetilde{\rho}_{V}(t)$
from \eqref{eq:exact-rV} drops
monotonically to the value $-\Lambda$, while the
standard-matter-type component $\rho_{M}(t)$ first
increases but then drops to zero due to
the expansion of the Universe.
Note that, for the simplest possible description of
the instability of de Sitter-spacetime, it appears necessary to
have at least two types of particle-production
energy densities, distinguished by their equation of state.

\section*{Acknowledgment}
\vspace*{-0mm}\noindent
It is a pleasure to thank A.A. Starobinsky and G.E. Volovik
for valuable comments.



\begin{thebibliography}{99}

\bibitem{Polyakov2007}
A.M. Polyakov,
\textit{Nucl. Phys. B} \textbf{797}, 199 (2008),
arXiv:0709.2899.

\bibitem{Polyakov2009}
A.M. Polyakov,
\textit{Nucl. Phys. B} \textbf{834}, 316 (2010),
arXiv:0912.5503.

\bibitem{KrotovPolyakov2010}
D. Krotov and A.M. Polyakov,
\textit{Nucl. Phys. B} \textbf{849}, 410 (2011),
arXiv:1012.2107.

\bibitem{Akhmedov2011}
E.T.~Akhmedov,
\textit{JHEP} \textbf{1201}, 066 (2012),
arXiv:1110.2257.

\bibitem{ZeldovichStarobinsky1977}
Y.B. Zeldovich and A.A. Starobinsky,
\textit{JETP Lett.} \textbf{26}, 252 (1977).

\bibitem{Starobinsky1979}
A.A.~Starobinsky,
\textit{JETP Lett.}  \textbf{30}, 682 (1979).

\bibitem{HawkingEllis1973}
S.W.~Hawking and G.F.R.~Ellis,
\textit{The Large Scale Structure of Space-Time}
(Cambridge University Press, Cambridge, England, 1973).

\bibitem{GibbonsHawking1977}
G.W.~Gibbons and S.W.~Hawking,
\textit{Phys. Rev. D} \textbf{15}, 2738 (1977).

\bibitem{KlinkhamerVolovik2008a}
F.R. Klinkhamer and G.E. Volovik,
\textit{Phys. Rev. D} \textbf{77}, 085015 (2008),
arXiv:0711.3170.

\bibitem{KlinkhamerVolovik2008b}
F.R. Klinkhamer and G.E. Volovik,
\textit{Phys. Rev. D} \textbf{78}, 063528 (2008),
arXiv:0806.2805.

\bibitem{Wald1984}
R.M. Wald,
\emph{General Relativity}
(Chicago University Press, Chicago, USA, 1984).

\end{thebibliography}
\end{document}